# THE RF SOURCE OF THE 60-MEV LINAC FOR THE KEK/JAERI JOINT PROJECT

S. Fukuda, S. Anami, C. Kubota, M. Kawamura, S. Yamaguchi, M. Ono,
H. Nakanishi, KEK, Tsukuba, 305-0035 Japan
S. Miyake, M. Sakamoto, Electron Tube and Device Division, Toshiba, Ohtawara, 324-8550 Japan,

*Abstract*

The construction of the 60-MeV proton linac has started as a low-energy front of the KEK/JAERI Joint Project for a high-intensity proton accelerator facility at KEK. The accelerating frequency is 324 MHz. Five UHF klystrons are used as an rf source; their ratings have a maximum power of 3 MW, a beam pulse width of a 700 μsec (an rf pulse width is 650 μsec) and a repetition rate of 50 pps. We have manufactured a proto-type rf source (a power-supply system with a modulating-anode pulse modulator and prototype klystrons). In this paper, the specifications and developments of the rf source, including the WR-2300 waveguide system, are summarized. During the manufacturing process, strong oscillations due to backward-moving electrons from the collector were observed. This phenomenon was analyzed both experimentally and theoretically. We have tested up to an output power of nearly 3 MW, and succeeded to test the DTL hot-model structure.

## 1 INTRODUCTION

The KEK/JAERI Joint Project, a currently proposed accelerator complex, comprises a 600-MeV proton linac, a 3-GeV rapid-cycling synchrotron and a 50-GeV synchrotron [1]. The 600-MeV linac comprises a 200-MeV, 324-MHz low-β section, a 972-MHz section from 200 MeV to 400 MeV, and a 972-MHz super-conducting section from 400 MeV to 600 MeV. The construction of the 60-MeV linac in KEK has been started as a low-energy front of this KEK/JAERI Joint Project. The 60-MeV linac comprises a negative hydrogen ion source, a 3-MeV RFQ linac, a 50-MeV DTL and a SDTL [2]. The accelerating frequency is 324 MHz. High-power and high-duty rf-sources are required for these structures, and a 324-MHz klystron with a modulating-anode has been adopted as the rf-power source. Although the maximum rating of the klystron is 3 MW for 650 μsec rf pulse duration and a 50 pps repetition rate, the actual working level of the klystron is 2 MW. Because this working rf-power must be well controlled both in amplitude and phase, the klystron is required to generate a saturated output power of 2.5 MW at the same duty and be operated at the 2-MW output power level by controlling the driving power. Since a klystron with a modulating-anode is used, the high-voltage power supply at the test station consists of an old JHF dc-cathode voltage supply and a newly developed pulse-modulating anode voltage supply. Newly developed power supplies [3] are now installed in the new building. Since a 324-MHz klystron is the lowest-frequency one in practical use, we manufactured a high-power test model of a co-axial window [4] and a high-power beam test tube composed of an electric gun and a collector [5] to confirm the technical feasibility. Prototype klystrons were manufactured and tested from 1999 and high-power tests of the klystrons, the hot model of a DTL structure and WR-2300 waveguide components were successfully performed. A low-power system of the klystron (the driving system) was also developed. This low-power line is required for a precise feedback system for the amplitude and the phase control to establish operation at the unsaturated output power region by controlling the driving power.

Table 1: Specifications of the 324-MHz klystron.

| Item | Unit | Max. | Working(Sat.) |
|---|---|---|---|
| Operating frequency | MHz | | 324 |
| Peak output power | MW | 3.0 | 2.5 |
| Beam pulse width | μs | | 700 |
| RF pulse width | μs | | 650 (flat top: 620) |
| Repetition rate | pps | | 50 |
| RF duty | % | | 3.25 |
| Beam current | A | 50 | 45 |
| Beam voltage | kV | 110 | 102 |
| Mod. anode voltage | kV | 93 | 86 |
| Micro-perveance | | | 1.37 |
| Efficiency | % | | 55 |
| Gain | dB | | 50 |
| Input / Output port | | | N-type / WR-2300 |
| RF window | | | coaxial ceramic window |
| Mounting position | | | horizontal |
| Focusing | | | electromagnet focusing |

## 2 ANODE MODULATOR FOR THE 324-MHZ KLYSTRON

Since the 324-MHz klystron was planed to be mounted horizontally with its oil tank socket, it was necessary to develop a new independent anode modulator. This was installed in a compact oil tank; one pair of filaments in the high-voltage cathode potential and a modulating anode connection are connected to the klystron socket. A DC high voltage was supplied from the old DC power supply to this anode modulator through a high-voltage cable in the test station. A hard-tube switching device (TH-5188) and associated G1, G2 power supplies were assembled as a module; further, it is easy to replace. This modulator produced a pulse with a rise time of 32 μsec and a fall time of 183 μsec. Using the test beam tube mentioned before, this modulator was operated up to the full rating of about 93 kV at a cathode voltage of 110 kV with a pulse width of 750 μsec and a repetition rate of 50 pps. Recently, a high-voltage transistor switch was developed to replace the hard tube, and was successfully tested at KEK.

## 3 DRIVER AND HIGH-POWER WAVEGUIDE SYSTEM

Although the driving system for the 324-MHz klystron in the test station simply comprised a signal generator, a pulse modulator, an attenuator and an amplifier, a feedback control system was recently constructed and tested in the low-power driving system [7]. This was done because the high precision control of the rf amplitude of less than 1% and a phase of less than 1 degree is required in order to mate the injection requirements to the 3-GeV synchrotron so that $\Delta p/p$ is less than 0.1%.

The WR-2300 rectangular waveguide system comprises a directional coupler, a 3-dB hybrid, a phase-shifter of the triple-stub type, a high-power circulator and a co-axial (WX-203D) to a rectangular-waveguide transition. Since the output power from a klystron is fed to two accelerating structures, the high-power WR2300 rectangular waveguide system must have a precise 3dB-hybrid power-divider and a phase shifter in one port after the power divider due to phase and amplitude requirements. A design refinement is now progressing. A high-power waveguide circulator was also developed in order to protect the klystron from any large reflection power at the rising time and the falling time of the pulse. These components have been successfully tested using the high-power klystron.

## 4 PERFORMANCE OF THE 324-MHZ KLYSTRON

### 4.1 Klystron Developments

The high-power klystron with a modulating anode was adopted as the rf source while considering the entire project scheme[4]. The specifications are given in Table 1. The operating frequency, 324 MHz, was lower than that of the CERN 1MW tube, which was the lowest-frequency klystron practically used in a large-scale accelerator; therefore, as a first step, a high-power model of a co-axial window and a beam test tube were manufactured and tested in 1998. The former was tested up to 2-MW output power using the 432-MHz JHP test facility. The beam test tube was operated up to 110-kV dc cathode voltage, 720 μsec beam pulse width and 50 pps repetition [5]. The prototype klystrons were then developed and tested from 1999 to 2000 and used for high-power tests of the waveguide components and accelerator structure. This klystron is 4.5m long, from the bottom of the oil tank to the top of the collector, and was mounted horizontally. Figure 1 shows a picture of the test station at KEK.

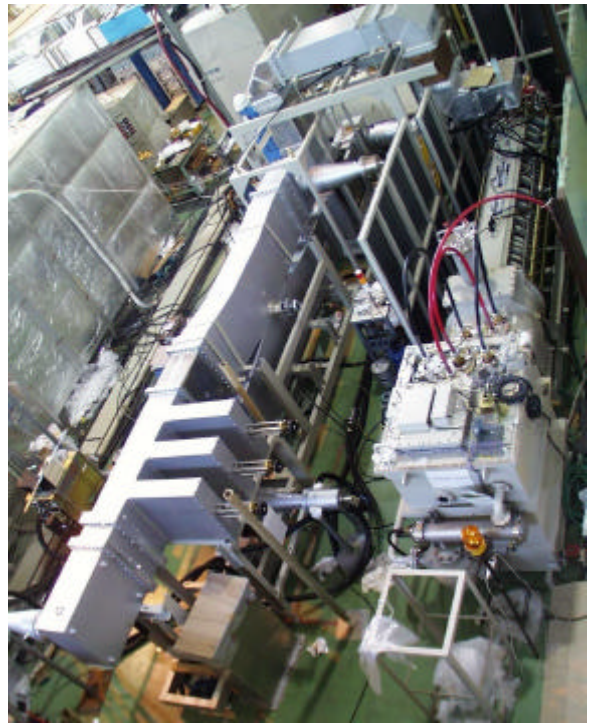

Figure 1: High-power test of the klystron at KEK.

During the first test, strong spurious oscillations were observed. These occurred under high-voltage operation in the range of 65-72 kV and higher than 90 kV without any driving input power. The operating frequency was nearly the same as the operating one. These oscillations were concluded to be caused by backward-moving electrons from the collector, since the some of the oscillations ceased when a weak deflecting magnetic field was applied in the collector region. Since this klystron's operating frequency is low, the drift-tube radius is large and the aspect ratio of diameters of a drift-tube to a collector is small compared with the other frequency-band klystron. From a simulation using the EGS4 code, it became clear that the backward-moving electrons could be decreased by using a larger diameter collector[6]. Several experiments with different collector shapes were performed to eliminate these oscillations and the associated unstable phenomena with an input drive power. In order to investigate the output power characteristics, the Rieke diagram was measured by changing the reflection from the load using a triple-stub-type phase shifter as the reflection device. This was also useful to optimize the klystron circuit parameters.

*4.2 Characteristics of the Klystron*

So far, three klystrons were built and tested at KEK. After investigating the backward-moving electrons from the collector, the spurious oscillation was successfully eliminated and the klystron data were measured. The latest test was performed up to the 107-kV cathode voltage. A limitation came from arcing of the switching tetrode in the modulating anode power supply. Figure 2 shows the output-power characteristics with the function of the applied voltage. Figure 3 shows the output characteristic with the function of the input drive power (Figure 3 left) and that with the function of the frequency of the input drive signal (Figure 3 right).

Nearly 3MW output power and the efficiency of 52% were obtained when the anode voltage was applied at a value 10% higher than the nominal operating condition at a cathode voltage of 106.6kV. Lower-gain operation was preferable for more stable output characteristics. For more stable operation and higher efficiency, further design development might be considered.

## 5 SUMMARY

A test of the rf source of the 60-MeV linac was successfully performed at KEK. The 324-MHz high-power klystron exceeded the saturated output power of 2.5 MW, the required power of the working point. The power supply and the anode modulator were also operated up to their full ratings. High-power waveguide components and an accelerator structure test are under development. From the fall of 2000, construction of the 60-MeV linac for the KEK/JAERI Joint Project will be conducted in a new building.

## REFERENCES

[1] Y. Yamazaki et al., "Accelerator Complex for the Joint Project of JAERI/NSP and KEK/JHF", 12th Symp. on Acc. Sci. and Tech., Wako, Japan, 1999.
[2] Y. Yamazaki et al., "The Construction of the Low-Energy Front 60-MeV Linac for the JAERI/KEK Joint Project", presented in this Conference.
[3] M. Ono et al., "Power Supply System for 324 MHz Klystron of the JHF Proton Linear Accelerator", 12th Symp. on Acc. Sci. and Tech., Wako, Japan, 1999.
[4] S. Fukuda et al., "Development of a High-Power VHF Klystron for JHF", APAC'98, Tsukuba, Ibaraki, Japan, 1998.
[5] M. Kawamura et al., "High-power test of a klystron beam-test-tube and an anode modulator" (in Japanese), the 24th Linear Acc. Meeting in Japan, Sapporo, Hokkaido, Japan, 1999.
[6] Z. Fang et al., "Simulation of Back-going Electrons from a Collector of Klystron", presented in this Conference.
[7] S. Yamaguchi et al., "Feedback Control for 324 MHz Klystron"(in Japanese), the 25th Linear Acc. Meeting in Japan, Himeji, Hyogo, Japan, 2000.

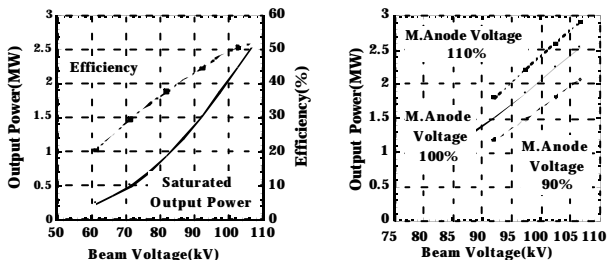

Figure 2: Output-power characteristics along with the function of the beam voltage.

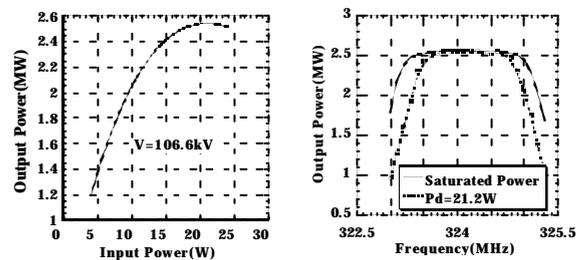

Figure 3: Output power characteristics with the function of the input power (left) and with the drive frequency (right).